\documentclass[baaa]{baaa}

\usepackage[pdftex]{hyperref}
\usepackage{subfigure}
\usepackage{natbib}
\usepackage{helvet,soul}
\usepackage[font=small]{caption}

\contriblanguage{1}

\contribtype{1}

\thematicarea{11}

\title{Fractal dimension in star formation regions}

\titlerunning{Fractal dimension in star formation regions}

\author{T. Canavesi\inst{1,2,3} \& S. Hurtado\inst{2,3}}

\authorrunning{Canavesi et al.}

\contact{tobiascanavesi@gmail.com}

\institute{ Instituto de Física de La Plata, CONICET-UNLP, Argentina \and Facultad de Ciencias Astron\'omicas y 
Geof\'isicas, UNLP, Argentina \and Consejo Nacional de Investigaciones Cient\'ificas y T\'ecnicas, Argentina}

\resumen{
	Varios estudios en dos dimensiones muestran que las zonas de formación estelar HII en diferentes galaxias del tipo espiral tienen una 
	dimensión fractal de aproximadamente $2.3$. En este trabajo se calcula la dimensión fractal a través del método de box counting mediante la implementación de un código en R. Lo innovador del trabajo 
	consiste en calcular la dimensión fractal directamente en 3 dimensiones, sin necesidad del uso de proyecciones, y 
	la utilización de las posiciones y distancias de la misi\'on \textsl{Gaia} (Data Release 2). La dimensión fractal estimada de las regiones estudiadas en la vía láctea es 2.468 (M16), 2.126 (Nebulosa de Orión) y 2.435 (RCW 38).
}

\abstract{
	Several two-dimensional studies in spiral galaxies show that HII star formation regions have a fractal distribution, with a fractal dimension of approximately 2.3. In this work, the fractal dimension is calculated through 
	the box-counting method implemented in an R code. The innovation of the work lies in calculating the fractal dimension
	directly in 3 dimensions, without the need for the use of projections, using position and distance data from 
	\textsl{Gaia} (Data version 2). The estimated fractal dimension for the regions studied in the Milky Way are 2.468 (M16), 
	2.126 (Orion Nebula) and 2.435 (RCW 38).
}


\begin{document}
	
	\maketitle
	
	\section{Introduction}
	\label{S_intro}
	Historically, interest in geometry has been stimulated by its applications to nature. The ellipse assumed importance as the shape of planetary orbits, as did the sphere as the shape of the Earth. Geometry of ellipse and sphere can be applied to these physical situations. A similar situation pertains to fractals. Recent physics literature shows a variety of natural objects that are described as fractals, cloud boundaries, topographical surfaces, coastlines, turbulence in fluids, and so on. None of these are actual fractals, their fractal features disappear if they are viewed at sufficiently small scales. Nevertheless, over certain ranges of scale they appear very much like fractals, and at such scales may usefully be regarded as such.\\
	Several works suggest a fractal structure at different scales of the universe such as galaxy clusters or HII regions \citep{1987PhyA..144..257P,2001AJ....121.1507E,sanchez2010fractal}. Fractal objects share a symmetry called scale invariance, so they are invariant under a transformation which change a small part of a picture for a bigger one. We can characterize fractal systems by calculating their fractal dimension (FD).  \\
	In Sec.~\ref{data}, we introduce the data of the three star formation regions that have been studied. Then in Sec.~\ref{box}, the box-counting dimension is explained as a way to estimate the FD for a data set, and the computed FD is given for each region. Finally, we present the conclusions and the importance of a good algorithm for FD estimation (see Sec.~\ref{conclusions}).\\
	
	\section{Data}\label{data}
	
	\textsl{Gaia} is a mission to chart a three-dimensional map of our Galaxy, \citep{gaia2016gaia}. It serves, among other things, to study the composition, formation and evolution of 
	the Galaxy. \textsl{Gaia} provides radial velocity and position measurements with the necessary precision to 
	produce data that will allow the study of more than one billion stars in our Galaxy and in the entire Local Group. 
	\\
	The Data Release 2 of the \textsl{Gaia} mission \citep{gaia2018gaia} was used to obtain the Cartesian coordinates 
	of stars for three star formation regions RCW 38, Orion Nebula and M16, which distances are directly taken from 
	\citet{2018AJ....156...58B}. With this, a spatial graph of the stars can be achieved and the box-counting dimension 
	computed (see Sec.~\ref{box}). For RCW 38 we consider a distance of 1700$~\rm{pc}$ with an error of $\pm 300~\rm{pc}$, and a 
	cube of	$300 \times 300~\rm{pc}$, for M16 a distance of 1750$~\rm{pc}$ with an error of  $\pm 300~\rm{pc}$, so we consider a cube of 
	$300 \times 300~\rm{pc}$, and for Orion Nebula a distance of $400$$~\rm{pc}$ with an error of $\pm 100~\rm{pc}$, so consider a cube of
	cube of $100 \times 100~\rm{pc}$.
	
	\begin{figure}[h]
		\centering
		\includegraphics[width=0.4\textwidth]{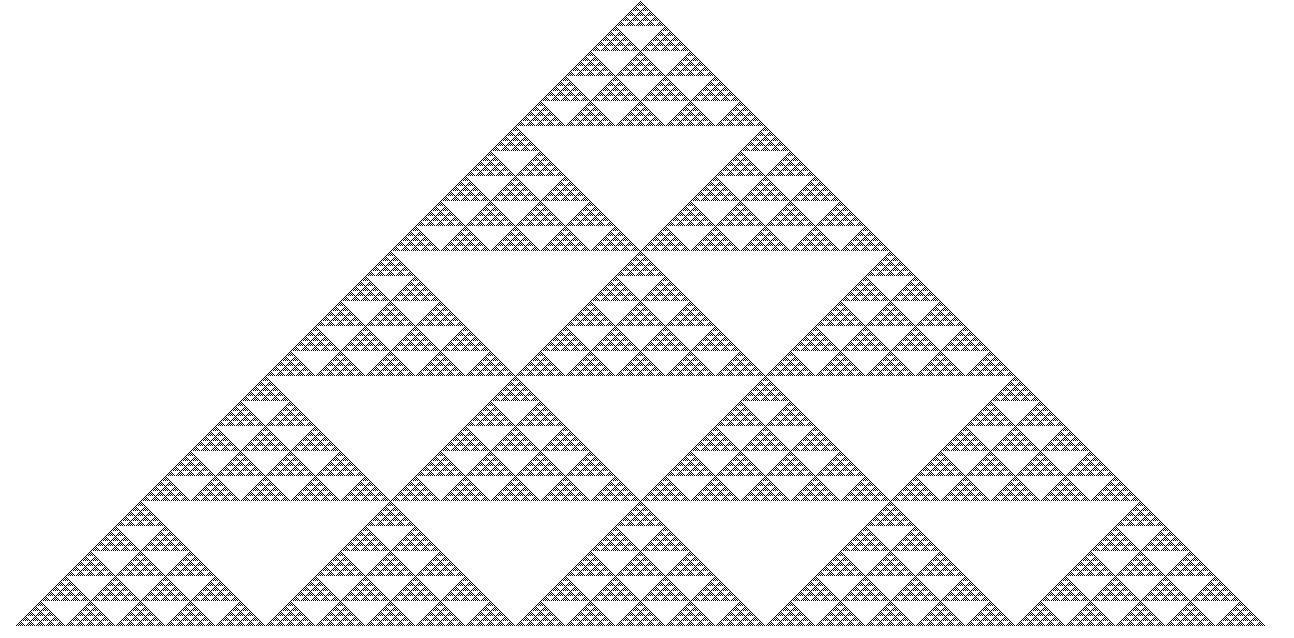}
		\includegraphics[width=0.4\textwidth]{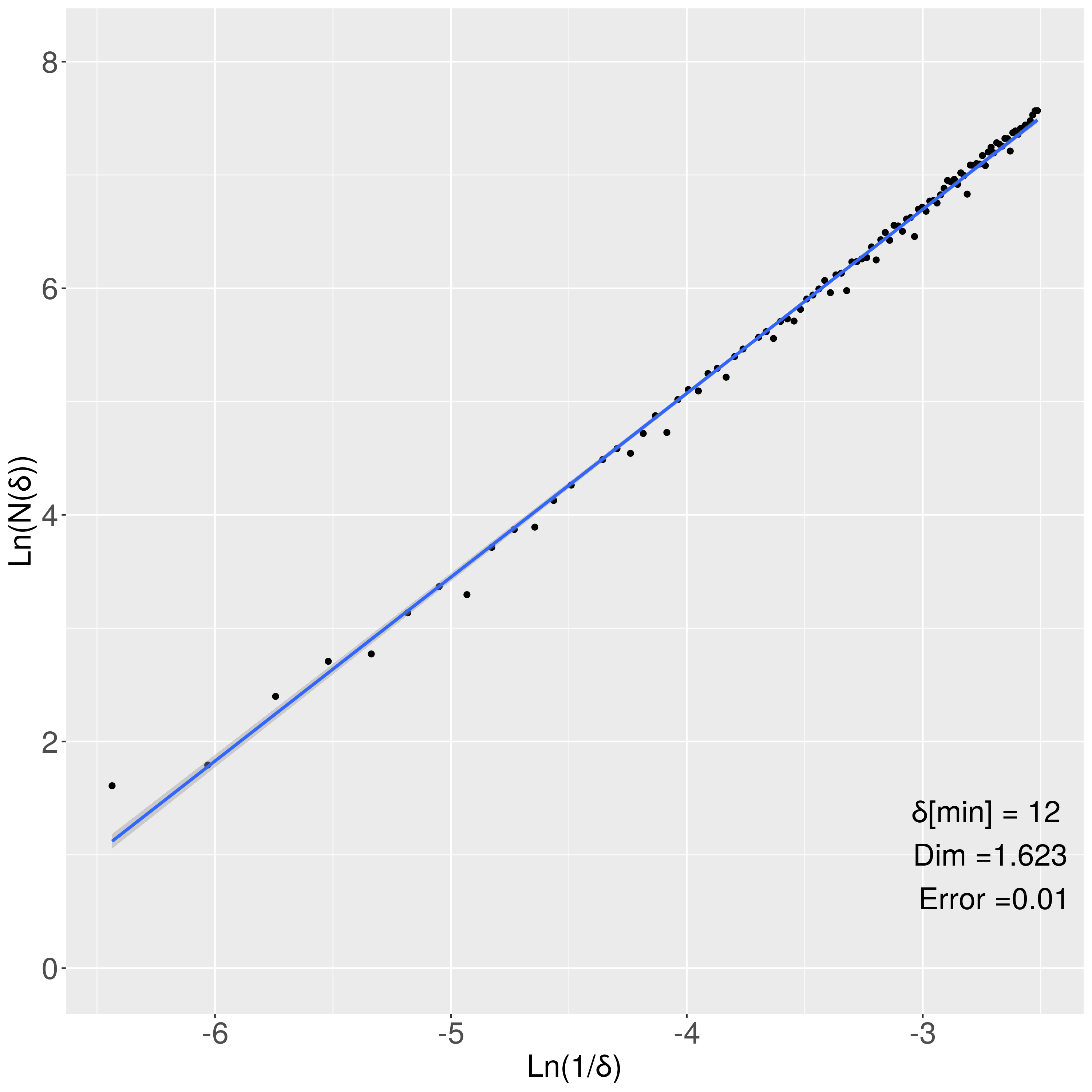}
		\caption{Top panel: Pascal triangle module 3. Bottom Panel: fitted FD of the  Pascal triangle module 3.}
		\label{triangle}
	\end{figure}
	
	\begin{figure*}[]
		\centering
		\includegraphics[width=0.3\textwidth]{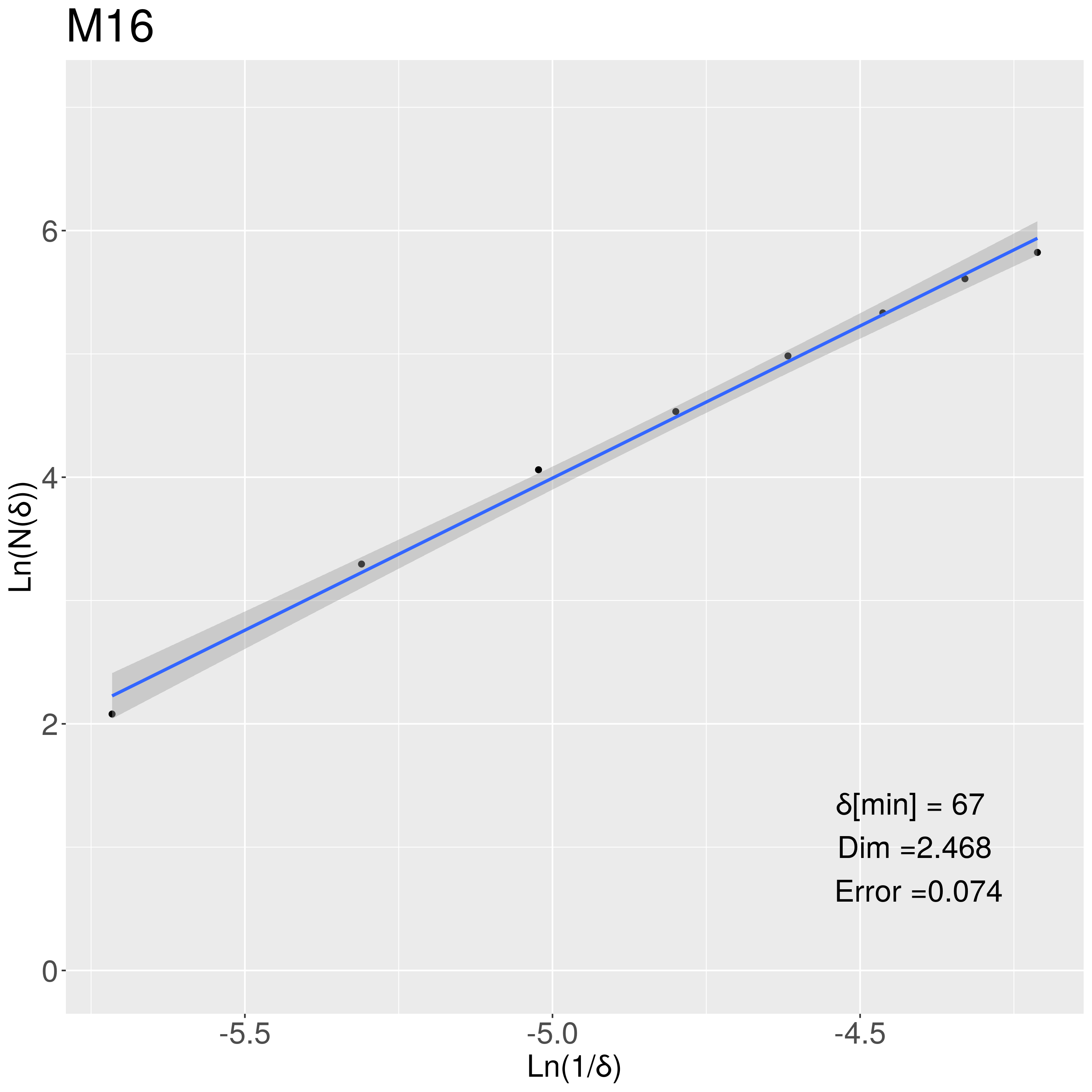}
		\includegraphics[width=0.3\textwidth]{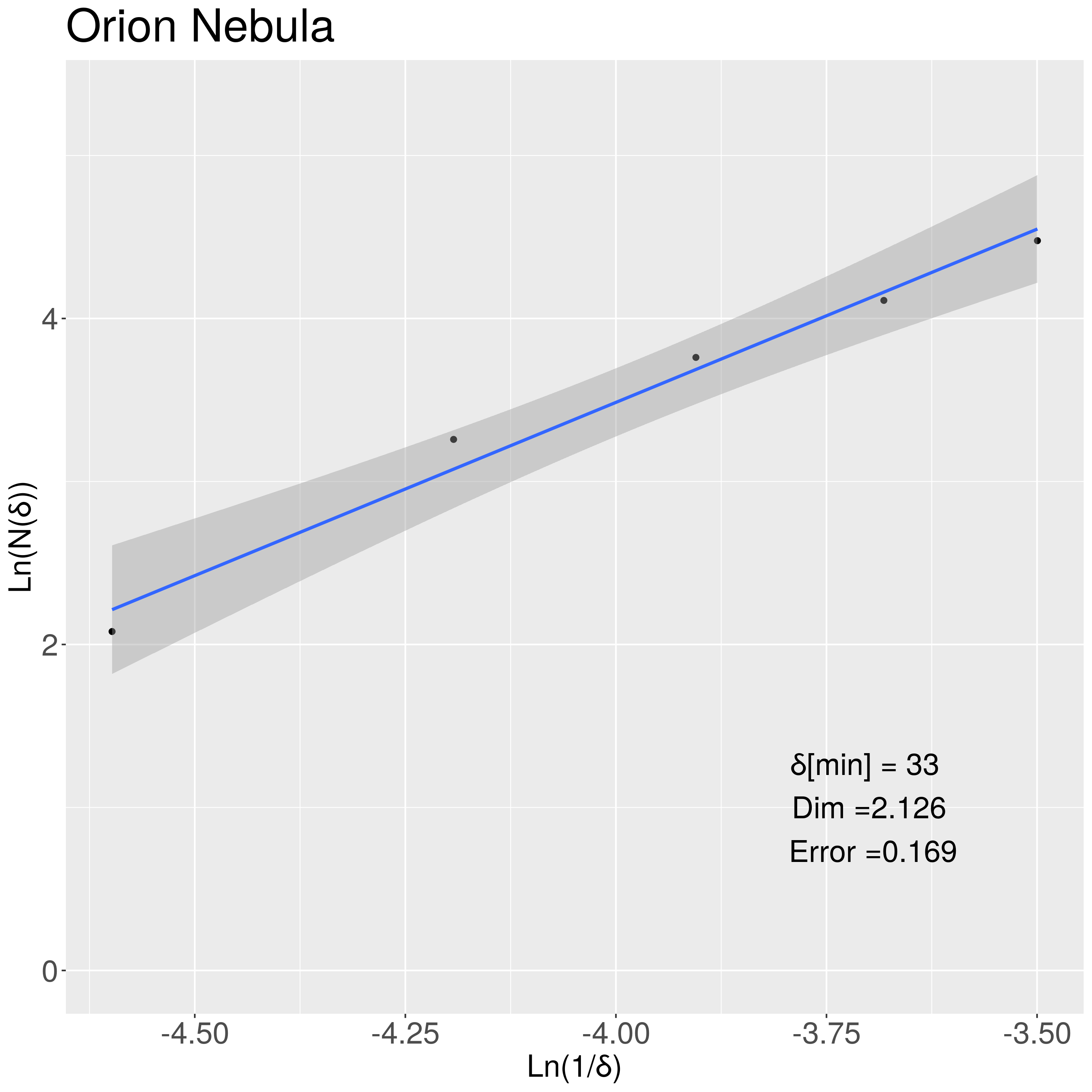}
		\includegraphics[width=0.3\textwidth]{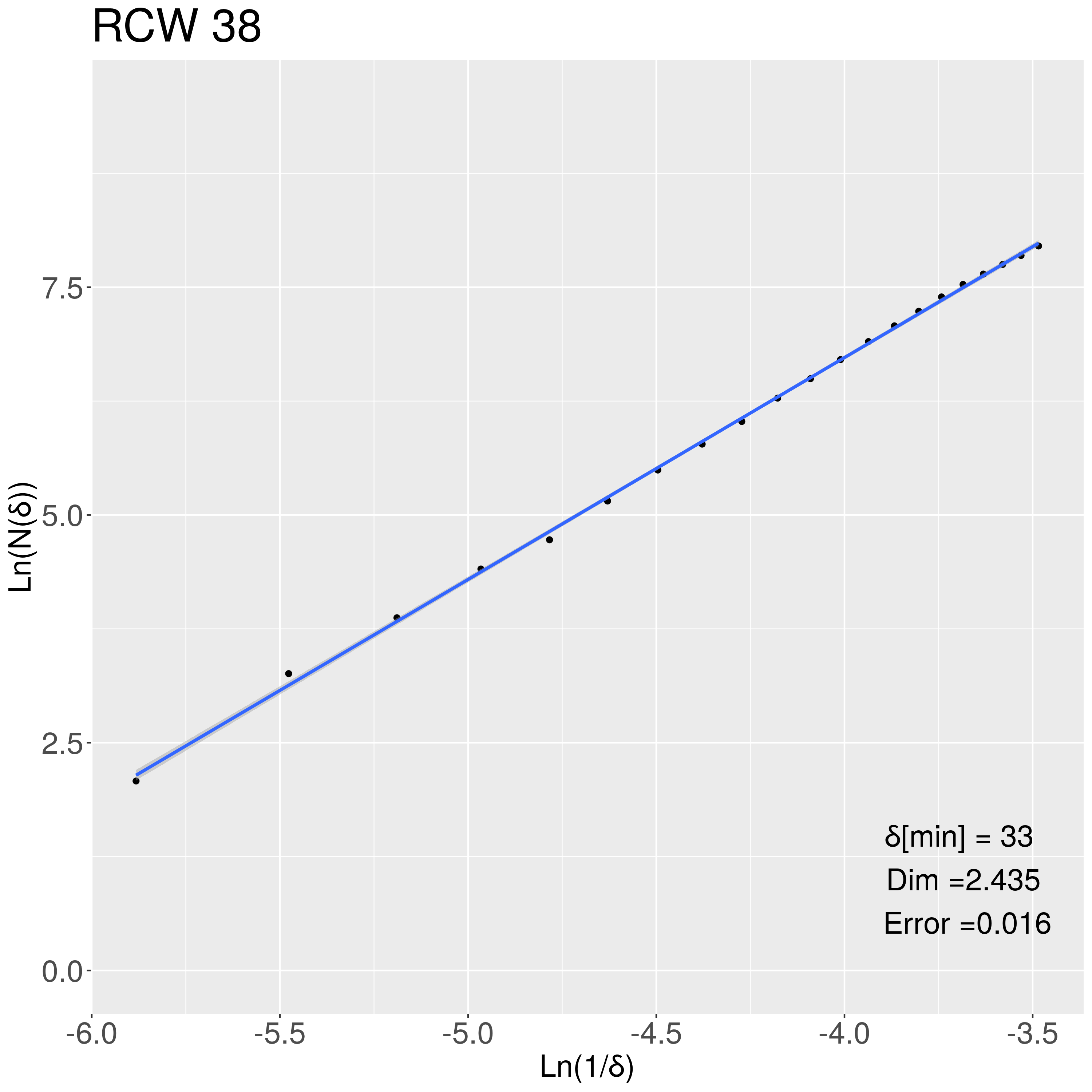}
		\caption{ FD fitted for the three studied regions, where $ \delta[min]$ is the smallest length scale in pc and 
		Dim is the FD (see Eq.~\ref{eq1}).}
		\label{ajustes}
	\end{figure*}
	
	\section{Box-counting Dimension}\label{box}
	A fractal is by definition a set for which the Hausdorff-Besicovitch dimension strictly is not equal to the 
	euclidean dimension. The Hausdorff-Besicovitch dimension is not practical to compute so alternative definitions are 
	used. The most common and used definition is the Box-Counting dimension \citep{feder2013fractals}. Let be $F$ a 
	subset $F \in \mathbf{R}^n$, with $N_\delta(F)$ the number of $\delta-mesh$ cubes that intersect $F$, the 
	boxcounting dimension ($dim_b$) is:
	
	\begin{equation}
	\label{eq1}
	dim_b F = \lim_{\delta\to 0} \frac{\log N_\delta (F)}{-\log \delta}.
	\end{equation}
	
	Note that $dim_b F$ is the slope of $\log N_\delta (F)$ vs. $-\log \delta$.
	We developed the algorithm to calculate this dimension in 2D and 3D in R ~\citep{rrr}, avoiding both boundary and small data set problems. In order to test the algorithm, we compute the fractal dimension of a known fractal. The used fractal was the Pascal triangle module 3 for which we obtained a $FD=1.623$ while the real FD is $1.6309$ (Fig.~\ref{triangle}).\\
	The definition of the Hausdorff dimension of a set of particles requires the diameter of the covering sets to 
	vanish. In our case of study, we have a characteristic smallest length scale, this is where the fractal structure 
	of HII regions disappears. We associate this value with $\delta_{min}$ which considering the three regions gives us 
	an average of $44.33~\rm{pc}$. To find the  $\delta_{min}$ we analyze the $\log_{10} N_\delta (F)$ vs. $-\log_{10} 
	\delta$ plot for all the stars and analyze when we lose the linear behavior.\\
	If we assume unchanged density, we can consider the fractal dimension as the mass dimension, this is based on the 
	idea of how a system mass scales, so:
	
	\begin{equation} \label{eq2}
	M(R) \sim R^{dim_b} \text{ ; } \delta_{min} \le \delta \le \delta_{max}.
	\end{equation}
	\vspace{1mm}
	Where $\delta_{max}$ is the size of the studied domain. If we assumed that matter with constant density is 
	distributed over the fractal, then the mass of the fractal enclosed in a volume of a characteristic size R 
	satisfies the power law of Eq.~\ref{eq2}, with $dim_b \notin \mathbb{Z}$. Whereas for a regular $3D$ Eucliadian 
	object the dimension mass scales: $M(R) \sim R^3$. For more details see \citet{terasov2010}. 
	
	\begin{table}[h]
		\centering
		\small
		\caption{ Coordinates and FD of the three studied regions.}
		\begin{tabular}{p{11mm}p{14mm}p{16mm}p{6mm}c}
			\hline\hline\noalign{\smallskip}
			\!\!\!\!Object &\!\!\!\! $\alpha$ &\!\!\!\! $\delta$ &\!\!\!\!Distance &\!\!\!\! FD \\
			\!\! &\!\!\!\!  (J2000) &\!\!\!\! (J2000) &\!\!\!\! (pcs) &\!\!\!\! \\
			\hline\noalign{\smallskip}
			\!\!\!\!M16  & 18:18:48 & -13:48:24 & 1750 & 2.468$\pm$0.074\\
			\!\!\!\!Orion Nebula& 05:35:17.30 & -05:23:28 & 400 & 2.126$\pm$0.169\\
			\!\!\!\!RCW 38& 08:59:05.50  & -47:30:39.4& 1700 & 2.435$\pm$0.016\\	
			\hline
			\label{ajustes2}
		\end{tabular}
	\end{table}
	
Through our box-counting algorithm, we were able to obtain the FD of three star formation regions in the Milky Way 
using \textsl{Gaia} data. The obtained FDs are: $FD_{M16}=2.468$, $FD_{OrionNebula}=2.126$ and $FD_{RCW38}=2.435$ (see 
Tab.~\ref{ajustes2} and Fig.~\ref{ajustes}). These values are in concordance with FD values ($FD \sim 2.3$) obtained by 
\citet{2001AJ....121.1507E} for stars formation regions in different galaxies.\\
	
\section{Conclusions}\label{conclusions}
We found that the three star formation regions have a fractal dimension near $2.3$. This is important to support that 
the mass-size relationship can result from the fractal nature of this type of regions. Note that, these results are 
independent of the star formation distance, as was previously argued by \citet{Elmegreen2009}.  \\
The innovative aspect of this work is that the FD is calculated for the first time using \textsl{Gaia} data in the 
Milky Way, without using projections as necessary for two dimensions, achieving values similar to those obtained by 
\citet{2001AJ....121.1507E}. \textsl{Gaia} release can be used to further studied the fractal properties of the Milky 
Way.


	
	\bibliographystyle{baaa}
	\small
	\bibliography{bibliografia}

\begin{thebibliography}{10}
\providecommand{\natexlab}[1]{#1}

\bibitem[{{Bailer-Jones} et~al.(2018)}]{2018AJ....156...58B}
{Bailer-Jones} C.A.L., et~al., 2018, \aj, 156, 58

\bibitem[{Elmegreen \& Falgarone(2009)}]{Elmegreen2009}
Elmegreen B., Falgarone E., 2009, The Astrophysical Journal, 471, 816

\bibitem[{{Elmegreen} \& {Elmegreen}(2001)}]{2001AJ....121.1507E}
{Elmegreen} B.G., {Elmegreen} D.M., 2001, \aj, 121, 1507

\bibitem[{Feder(2013)}]{feder2013fractals}
Feder J., 2013, \textit{Fractals}, Springer Science \& Business Media

\bibitem[{{Gaia Collaboration} et~al.(2016)}]{gaia2016gaia}
{Gaia Collaboration}, et~al., 2016, \aap, 595, A1

\bibitem[{{Gaia Collaboration} et~al.(2018)}]{gaia2018gaia}
{Gaia Collaboration}, et~al., 2018, \aap, 616, A1

\bibitem[{{Pietronero}(1987)}]{1987PhyA..144..257P}
{Pietronero} L., 1987, Physica A Statistical Mechanics and its Applications,
  144, 257

\bibitem[{{R Core Team}(2019)}]{rrr}
{R Core Team}, 2019, \textit{R: A Language and Environment for Statistical
  Computing}, R Foundation for Statistical Computing, Vienna, Austria

\bibitem[{Sanchez \& Alfaro(2010)}]{sanchez2010fractal}
Sanchez N., Alfaro E.J., 2010, The fractal spatial distribution of stars in
  open clusters and stellar associations

\bibitem[{Tarasov(2011)}]{terasov2010}
Tarasov V.E., 2011, \textit{Fractional dynamics: applications of fractional
  calculus to dynamics of particles, fields and media}, Springer Science \&
  Business Media

\end{thebibliography}
	
\end{document}